\documentclass[final,3p,times,twocolumn]{elsarticle}
\usepackage{graphicx,epsf,float,amssymb,amsmath,makecell}
\usepackage[mathlines]{lineno}
\usepackage{xspace}
\newcommand{\fig}[1]{Fig.~\ref{#1}}
\newcommand{\eqnref}[1]{Eq.~\eqref{#1}}
\newcommand{\Fig}[1]{Figure~\ref{#1}}
\newcommand{\tab}[1]{Table~\ref{#1}}

\newcommand{\genunit}[2]{\ensuremath{#1~\text{#2}}\xspace}
\newcommand{\meters}[1] {\genunit{#1}{m}}
\newcommand{\minutes}[1]{\genunit{#1}{min}}
\newcommand{\inch}[1]   {\genunit{#1}{in}}
\newcommand{\hours}[1]  {\genunit{#1}{h}}

\newcommand{\degrees}[1]{\ensuremath{#1^{\mathrm{o}}}\xspace}

\newcommand{\Cftft}{\ensuremath{^{252}\text{Cf}}\xspace}
\newcommand{\Natt}{\ensuremath{^{22}\text{Na}}\xspace}
\newcommand{\Csots}{\ensuremath{^{137}\text{Cs}}\xspace}

\newcommand{\Pn}{\ensuremath{\hat{P}_{n}}\xspace}
\newcommand{\LN}{\ensuremath{\mathcal{L}_{n}}\xspace}
\newcommand{\LG}{\ensuremath{\mathcal{L}_{\gamma}}\xspace}
\newcommand{\LO}{\ensuremath{\mathcal{L}_{o}}\xspace}

\newcommand{\LNxE}{\ensuremath{L_{n}(x;E)}\xspace}
\newcommand{\LGxE}{\ensuremath{L_{\gamma}(x;E)}\xspace}
\newcommand{\LNpE}{\ensuremath{L_{n}(x_{p};p,E_{p})}\xspace}
\newcommand{\LGpE}{\ensuremath{L_{\gamma}(x_{p};p,E_{p})}\xspace}
\newcommand{\sigfrac}{\ensuremath{f_{s}}\xspace}
\newcommand{\NS}{\ensuremath{N_{S}}\xspace}
\newcommand{\NB}{\ensuremath{N_{B}}\xspace}
\newcommand{\phihat}{\ensuremath{\hat{\phi}}\xspace}

\journal{Nuclear Instruments and Methods}

\begin{document}
\begin{frontmatter}

\title{Source detection at 100 meter standoff with a time-encoded imaging system}

\author[sandia]{J. Brennan}
\author[sandia]{E. Brubaker}
\author[sandia]{M. Gerling}
\author[sandia]{P. Marleau}
\author[sandia,michigan]{M. Monterial}
\author[sandia]{A. Nowack\fnref{nowat1}}
\author[sandia]{P. Schuster\fnref{nowat2}}
\author[sandia]{B. Sturm}
\author[sandia]{M. Sweany\corref{cor1}}
\address[sandia]{Sandia National Laboratory, Livermore, CA 94550, USA}
\address[michigan]{University of Michigan, Ann Arbor, MI}
\cortext[cor1]{Corresponding author: msweany@sandia.gov}
\fntext[nowat1]{Currently at the University of Tennessee at Knoxville}
\fntext[nowat2]{Currently at the University of Michigan}

\begin{abstract}
We present the design, characterization, and testing of a laboratory prototype radiological search and localization system.  The system, 
based on time-encoded imaging, uses the attenuation signature of neutrons in time, induced by the geometrical 
layout and motion of the system.  We have demonstrated the ability to detect a \genunit{\sim 1}{mCi} \Cftft radiological source at \meters{100} standoff with 90\% detection efficiency and 10\% false positives against background in \minutes{12}.  This same detection efficiency is met at
\genunit{15}{s} for a \meters{40} standoff, and \genunit{1.2}{s} for a \meters{20} standoff.
\end{abstract}


\end{frontmatter}

\section{Introduction}
The detection and localization of radiological sources 
in various environments is an important nuclear security capability.
Some scenarios 
require quick localization of sources in highly cluttered 
background environments, and others may demand 
detection of sources over large areas. Because of their 
relatively low and isotropic natural background, ability to 
penetrate shielding, and long attenuation length in air 
(approximately \meters{100} at fission energies), fast neutrons are a strong 
candidate signature of illicit nuclear material.  
However, despite the relatively low background flux, 
variability caused by environmental factors such as weather 
conditions (pressure and humidity), geographic location 
(geomagnetic rigidity), local scattering sources, and even 
solar cycle, lead to a systematic uncertainty in the absolute 
neutron background rate \cite{davis, davis2}.  For example, the dominant factor in the 
time variation for a fixed location is the solar cycle, causing a 30\% variation
 \cite{ziegler}.  This variability ultimately limits the detection 
 sensitivity of gross counting detectors. 

Neutron imaging can reduce susceptibility to background variability,
but in the case of double scatter imagers \cite{gerling} the 
efficiency is low, and coded-aperture imagers \cite{hausladen} have
a limited field of view and poor imaging signal to background.  Both 
systems typically involve large numbers of detector/electronics channels 
that could impede fieldability and introduce systematic variability
due to, for example, differences in photodetector gain and overall detector light collection efficiency.  
While gain variation and and light collection efficiency in these systems can be calibrated to 
reduce systematic variability, the large number of channels adds a time and labor intensive 
calibration step in any measurement.

The time-encoded imaging (TEI) system described in this paper, however, has a 360-degree field of view, 
low channel count leading to reduced susceptibility to systematics, and does not require double scatters for
localization, resulting in improved efficiency.  Recently, we reported on a two-dimensional
fast neutron imager using time-encoded imaging (2D-TEI)
\cite{brennan1}.  That system was designed as a proof 
of principle for an alternative to coded-aperture imaging, with the 
distinction that, rather than modulating the radiation field in space 
and recording the modulation with position sensitive detectors, 
the field was modulated in time and recorded with a time 
sensitive detector.  The main systematic effects for such a 
system are those that induce a time modulation 
with the same rotational period as the detector rotation, of which there are few.  
Presented here is another system based on the TEI concept; 
targeting the application of radiological search at large 
standoff as opposed to high-resolution imaging yields a 
distinct detector system, which we call 1D-TEI.  Most notably, 
the system uses large detector cells to increase sensitivity;
the signal is modulated in only one dimension, providing 
localization in azimuth for sources near the horizon; and the modulation is 
accomplished by the detectors themselves, rather than by a separate mask, to reduce 
inactive detector mass.

Here we report on the performance of a 1D-TEI system for source 
detection at large standoff.  Although our system is capable of detecting both neutron
and gamma sources, we have designed it for neutron detection because neutron backgrounds 
are more reliably isotropic.  The system design and detector 
response are described in Sections \ref{sec:det} and \ref{sec:cal}, 
data analysis is presented in Section \ref{sec:ana}, and measurement 
results for sources at several different stand-off distances are presented in 
Section \ref{sec:data}.

\section{Time-encoded neutron imaging system}
\label{sec:det}
\begin{figure}[tbp]\centering
\includegraphics[width=0.8\columnwidth]{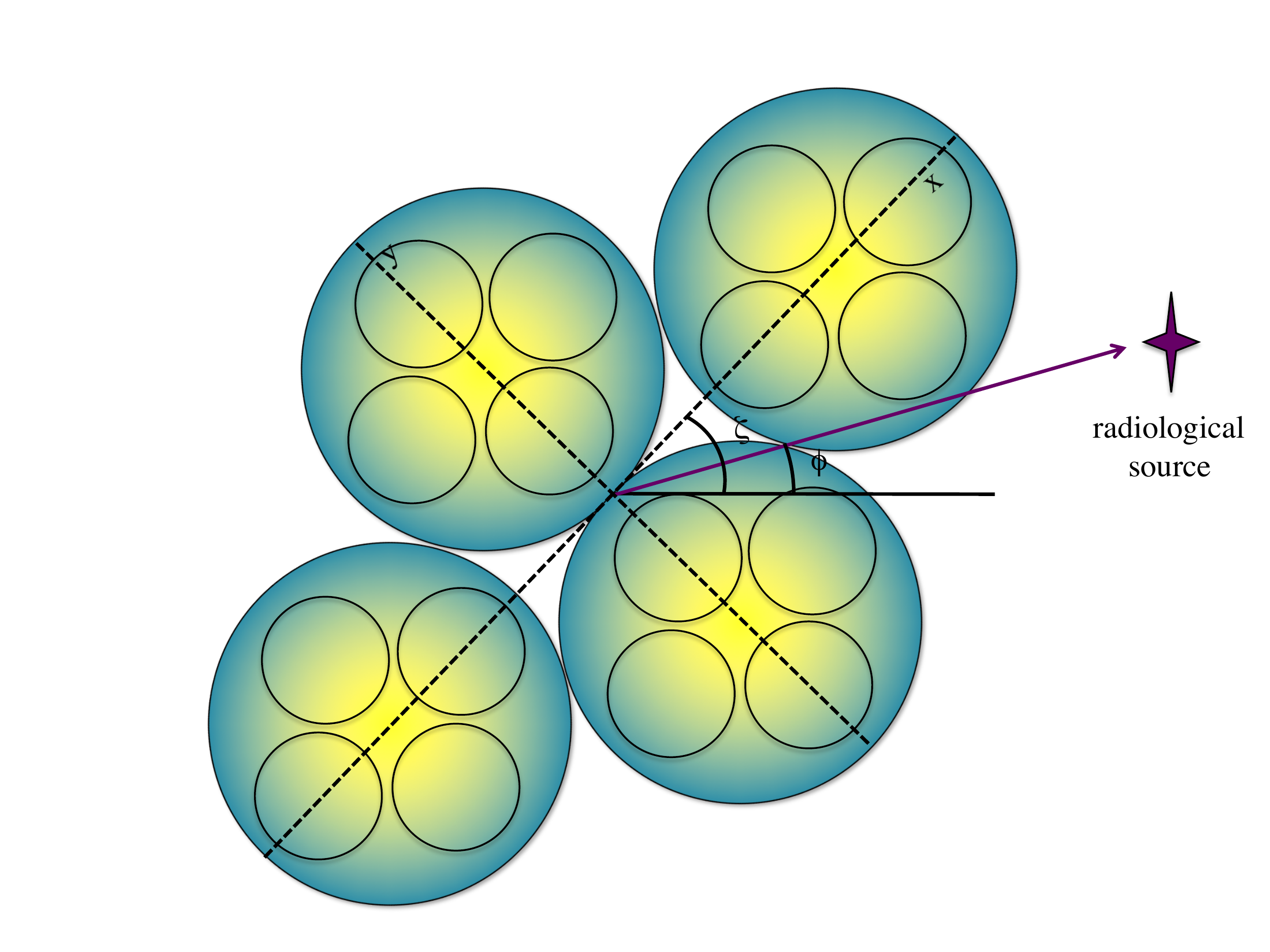}
\caption{A rendering of the one-dimensional time-encoded imaging system (1D-TEI).  
As the system is rotated by angle $\zeta$, radiation from the source, located at $\phi
$, is modulated by each of the detectors.  The PMT layout within the detector cells in 
indicated by the open black circles.}
\label{fig:fig1}
\end{figure}

\begin{figure}[tbp]\centering
\includegraphics[width=0.6\columnwidth]{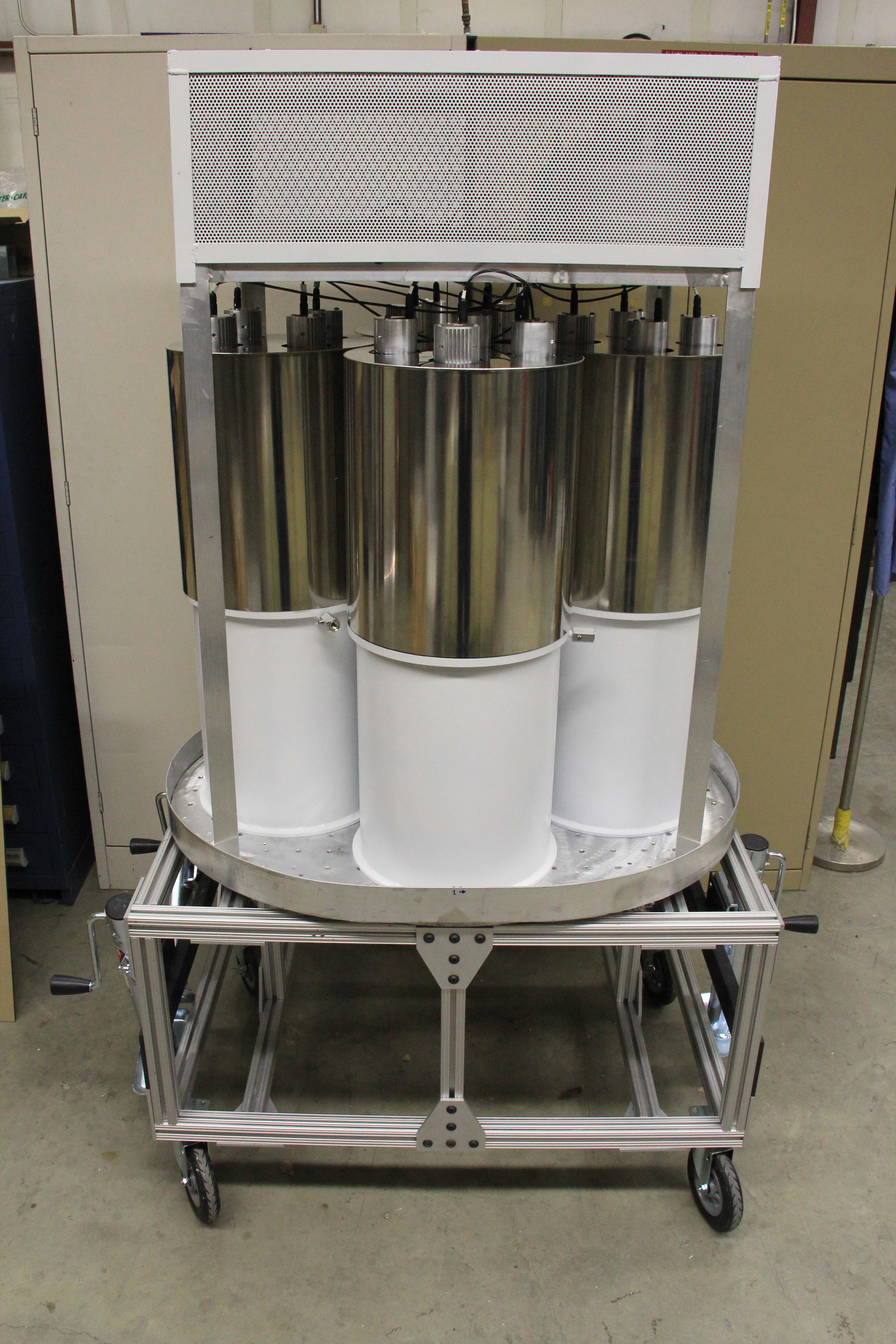}
\caption{Photo of the 1D-TEI instrument. The white-painted cells each contain
\genunit{\sim 27}{L} of EJ-309 liquid scintillator, and the aluminum cylinders
house four PMTs for each cell. The cells are arranged in a diamond pattern
on a rotational table, which is driven by a stepper motor.}
\label{fig:fig2}
\end{figure}

The 1D-TEI system consists of four neutron detector cells arranged in 
a diamond pattern that rotate around a common vertical axis.  Several different arrangements
of four detector cells were studied \cite{nowack2}, and this one was found to be optimal for source detection. As the 
system is rotated, the amount of shielding between each cell and a 
given radiological source location varies, modulating the fast neutron 
detection rate as a function of time.  The cell configuration is shown 
in \fig{fig:fig1}.  Liquid scintillator EJ-309 was used as a 
detection medium for its pulse shape discrimination (PSD) capabilities, 
safety benefits, and neutron attenuation properties.  The relevant design 
considerations were modulation and detection efficiency of fast neutrons.  
Considering modulation alone, larger detector cells are preferred for greater neutron 
attenuation, but pulse shape discrimination and therefore neutron detection 
efficiency is degraded with detector size due to decreased light collection efficiency and 
increased spread in their time of arrival.  We studied several configurations, detailed 
in \cite{nowack}, and found that $\inch{12}~\text{dia.} \times \inch{15}$ right cylindrical cells have acceptable 
PSD when read out by four 5-inch Hamamatsu H6527 photomultiplier tubes.  The 
liquid scintillator cells share an expansion reservoir with Argon gas overpressure and 
room to expand and contract within an estimated \degrees{80} F temperature swing. 
A picture of the 1D-TEI instrument is shown in \fig{fig:fig2}.

The liquid scintillator cells and electronics are all affixed to a turntable that is rotated 
by an Arcus stepper motor.  The motor is connected to a toothed wheel, and a
flywheel holds a geared belt against the inside bearing of the turntable which is also 
toothed: this ensures constant contact and prevents slipping.  A rotary encoder is 
attached to the turntable ring to measure the angle of rotation.  A metal divot 
attached to the turntable passes over a switch on the stationary mask frame to mark 
the end of one rotation.  During measurements, an Arcus controller board 
reads the encoder value, motor pulse value, and the state of the frame 
switch, which is written to hard disk once per rotation.  The 16 PMTs are read out by 
a Struck 3316 250 MHz desktop digitizer.  Each detector is independently and 
asynchronously triggered and read out by applying a threshold on the sum of the four 
PMT digital signals. In order to achieve continuous data acquisition, the data from 
the Arcus controller board and the Struck digitizer are read out by separate threads. 
The Struck digitizer has its own clock unit for recording time-stamps, but the Arcus 
controller relies on the acquisition computer CPU clock. The CPU clock and the 
Struck digitizer clock unit are synchronized at the beginning of the acquisition by 
simultaneously resetting the clock unit on the digitizer while sampling the CPU clock. 
We expect that any delay in executing these commands is on the order of 
milliseconds or less and thus inconsequential compared to the typical rotation rate of 
\genunit{0.5}{rpm}.

The amount of data from the Arcus-controlled board is relatively small and can be 
written to disk without any appreciable delay in acquisition.  The digitizer is read out 
whenever one of its two memory banks becomes full, at which point the acquisition 
switches to the other memory bank. If the data from the first bank can be 
downloaded to the acquisition computer and written to disk before the second 
memory bank fills up then the acquisition can continue without interruption.  The 
system can handle trigger rates of up to 360 kHz, after which events are lost. The event rates
for the data presented here are all at least an order of magnitude below this level.

\section{Detector Response}
\label{sec:cal}
There are three aspects of the detector response that must be calibrated to determine the neutron 
rate as a function of time.  First, a relative gain correction is necessary to combine the 
measurements for all PMTs on one cell to ensure azimuthal symmetry of the cell's response. However, a precise
absolute energy calibration for each cell is not required.  This is because the 
detection algorithm described below is robust against differences in efficiency between cells: it 
is the relative change in rate over time within each cell, due to the attenuation signature, that is critical to 
the detection algorithm.  Second, the pulse-shape discrimination must be characterized for 
each cell in order to distinguish gamma interactions from neutron interactions.  
Finally, the timing of each deposition must be correlated correctly with the rotational 
position of the turntable.

\subsection{Energy}
\begin{figure}[tbp]\centering
\includegraphics[width=0.9\columnwidth]{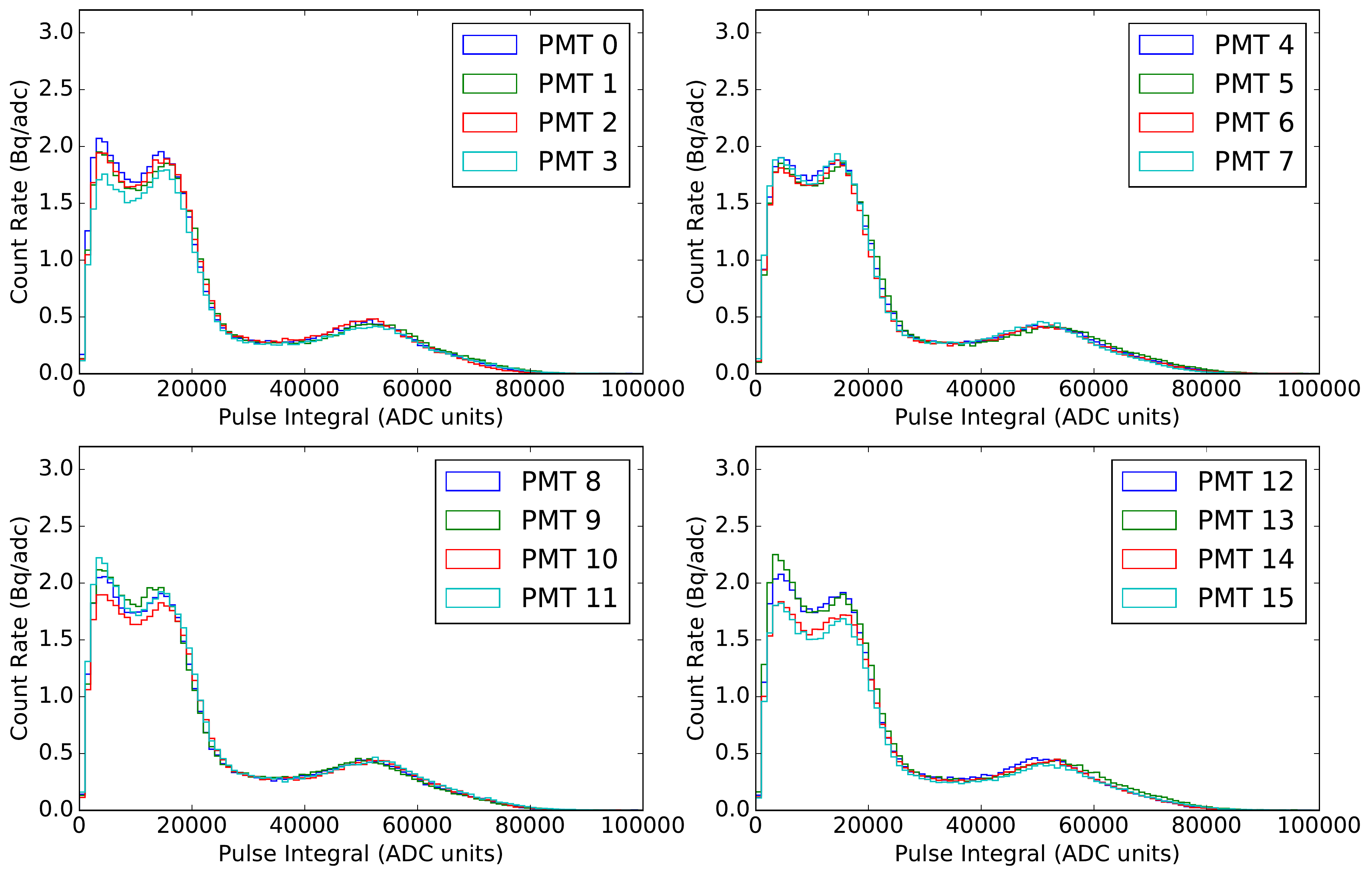}
\caption{The energy distribution resulting from a \Natt source for all 16 PMTs in the system.}
\label{fig:fig3}
\end{figure}

In order to equalize the relative response of the four detectors, we first gain-match 
the four individual photomultipliers in each detector.  This is accomplished by 
matching the normalized pulse-height spectrum resulting from a \Natt gamma-ray source,
centered below the cell so that it is equidistant from each PMT.  \Fig{fig:fig3} 
shows the distributions of the pulse integrals for all 16 PMTs in the system, indicating 
good overall agreement in both gain and efficiency.  The Compton edge from the 
1270 keV emission from \Natt is visible for all distributions between 50k and 
60k ADC units, and the Compton edge from the 511 keV positron annihilation 
gammas is visible between 10k and 20k ADC units.

\begin{figure}[tbp]\centering
\includegraphics[width=0.9\columnwidth]{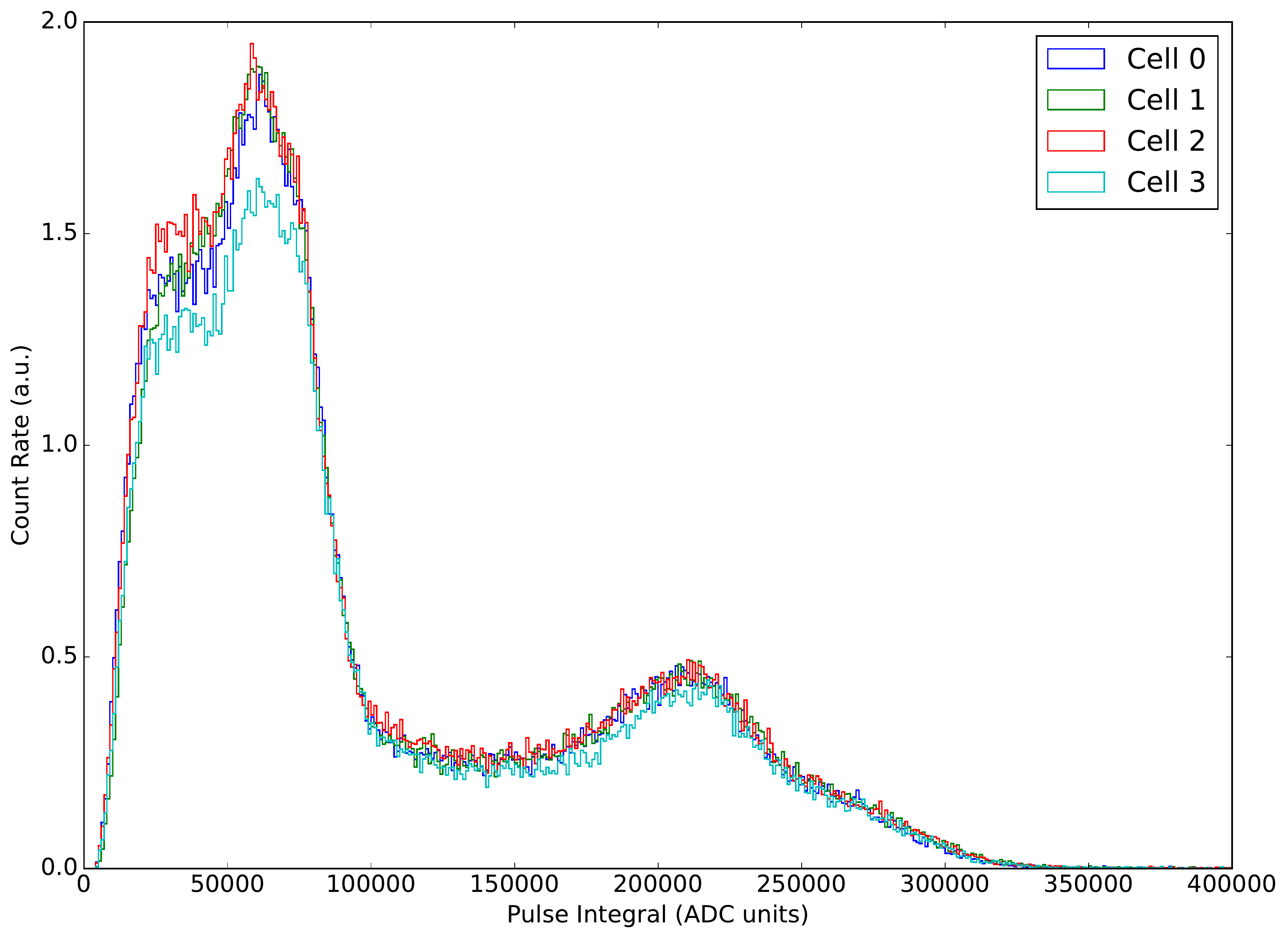}
\caption{The summed energy distribution resulting from a \Natt source for the four cells.}
\label{fig:fig4}
\end{figure}

Next, we reject events driven by spurious signals in one channel by adding a
requirement that all four PMTs individually exceed a noise threshold.
Using this selection, the integral of the sum of the four pulses is plotted in
\fig{fig:fig4}. Again, good overall agreement in gain is indicated by the alignment of the Compton 
features for each cell. Agreement in efficiency is indicated by similar event rates shown on the 
y-axis, with one cell slightly lower in rate than the other three: the total 
integrated event rates are within 10\% of each other, and the greatest difference is at lower energies.  As stated above, the 
detection algorithm is robust against small differences in efficiency from cell to cell.

Because this system depends on the modulation in rate as a function of 
rotation angle, it is necessary to confirm that there are no variations in efficiency or 
gain that are caused by factors other than the expected source attenuation.  
This was previously demonstrated for a similar photo-detector layout in the same 
size cell~\cite{nowack} by comparing the sum of photo-detector pulse heights from a 
\Csots source placed at the same height but different azimuthal angles around 
the cell.  No discernible difference was observed.  A difference in the response with 
height along the cell was observed, as expected due to the decrease in light 
collection as the average deposition height moves away from the PMT faces, 
however this does not affect the modulation of detection as a function of rotation 
angle.  

\subsection{Pulse-shape discrimination}
\begin{figure}[tbp]\centering
\includegraphics[width=0.9\columnwidth]{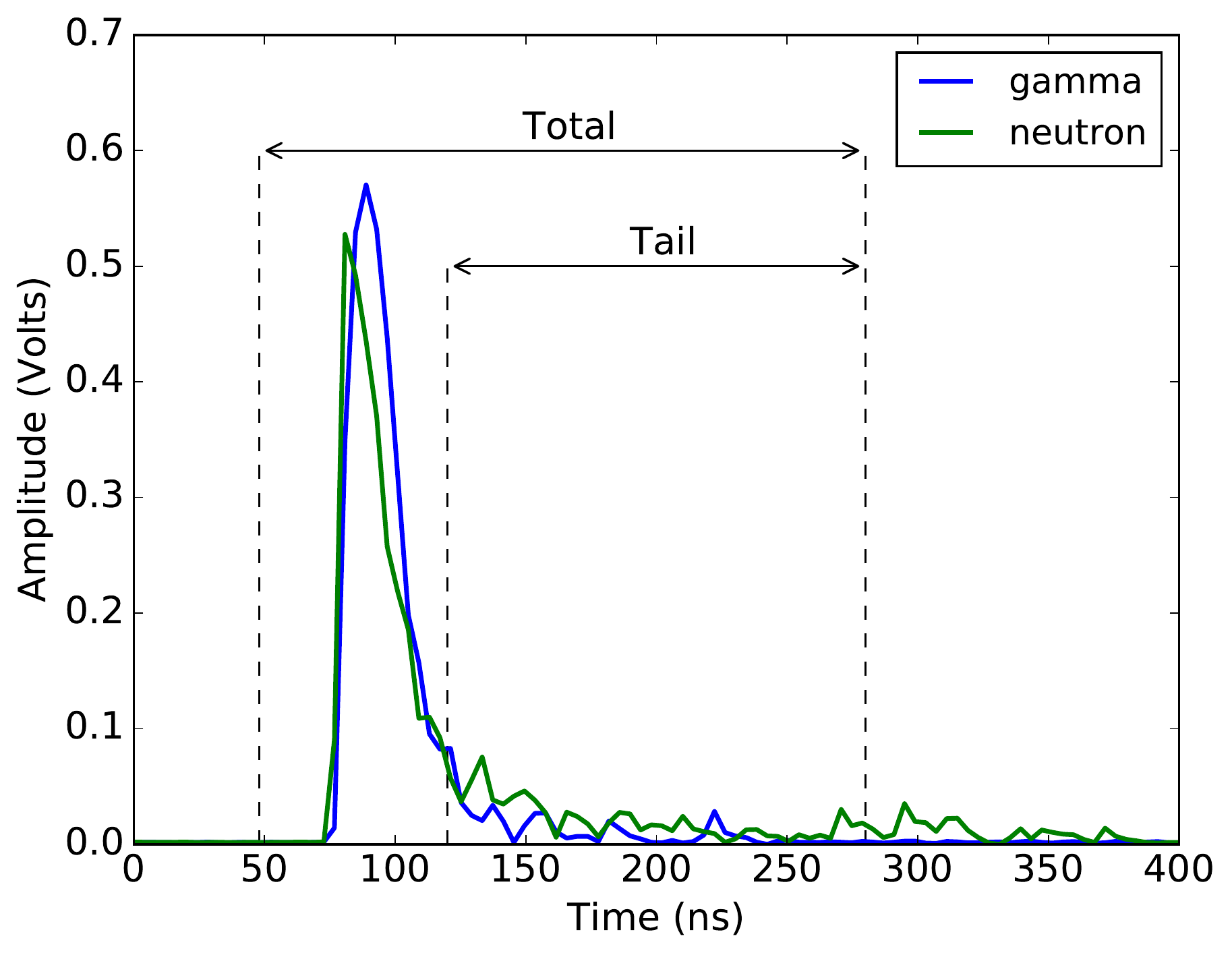}
\caption{An example neutron (green) and gamma (blue) pulse, with the gate lengths specified for the tail and total integrals.  The gates are determined by the trigger time of each pulse.  A window from 0 to \genunit{48}{ns} is used to determine the baseline.}
\label{fig:fig5}
\end{figure}

In order to discriminate between gamma and neutron depositions, we define a PSD 
parameter on the signal from each PMT as the ratio of the tail integral of the pulse
to a fixed total integral. The tail 
window was chosen to maximize the separation of the neutron and gamma distributions 
by optimizing a figure-of-merit (FOM), defined by:
\begin{linenomath}
\begin{equation}
\mathrm{FOM} = \frac{|\mu_n - \mu_\gamma|}{\Gamma_n + \Gamma_\gamma},
\end{equation}
\end{linenomath}
where $\mu_{n,\gamma}$ is the mean of the neutron, gamma distributions and
$\Gamma_{n,\gamma}$ is the width (FWHM).  The FOM is optimized for each cell, 
using the average PSD of the four PMTs vs. the summed pulse integral.  The tail 
and total windows, determined in reference to the trigger time of each pulse, are indicated in \fig{fig:fig5} for one of the cells. 
A baseline window is defined from 0 to \genunit{48}{ns}.

For the final event selection, 
we use a Bayesian approach to pulse-shape discrimination, similar to the method 
outlined in~\cite{monterial}.  Rather than averaging the PSD parameter values, the Bayesian
probabilities determined for each of the four PMTs are combined to produce one
overall probability for the cell.   This was demonstrated in \cite{monterial} to improve 
neutron/gamma discrimination compared to applying a decision boundary.
First, the PSD parameter distribution for each PMT is fit by a double Gaussian
distribution in slices of energy $E$ over the entire energy range. For a given slice: 
\begin{linenomath}
\begin{equation}
f(x,E) = \frac{1}{\sqrt{2\pi}} \left( \frac{A_\gamma(E)}{\sigma_\gamma(E)}e^{-\frac{(x-\mu_\gamma(E))^2}{2\sigma_\gamma(E)^2}} +  \frac{A_n(E)}{\sigma_n(E)}e^{-\frac{(x-\mu_n(E))^2}{2\sigma_n(E)^2}}\right).
\label{eq:dist}
\end{equation}
\end{linenomath}
where $x$ is the measured PSD value.  
We interpret the fitted Gaussians as likelihood functions:
\begin{linenomath}
\begin{align*}
\LGxE&=\frac{1}{\sigma_{\gamma}(E)\sqrt{2\pi}}e^{\frac{(x-\mu_{\gamma}(E))^2}{2\sigma_{\gamma}(E)^2}}; \\
\LNxE&=\frac{1}{\sigma_{n}(E)\sqrt{2\pi}}e^{\frac{(x-\mu_{n}(E))^2}{2\sigma_{n}(E)^2}}.
\end{align*}
\end{linenomath}

\begin{figure}[tbp]\centering
\includegraphics[width=0.9\columnwidth]{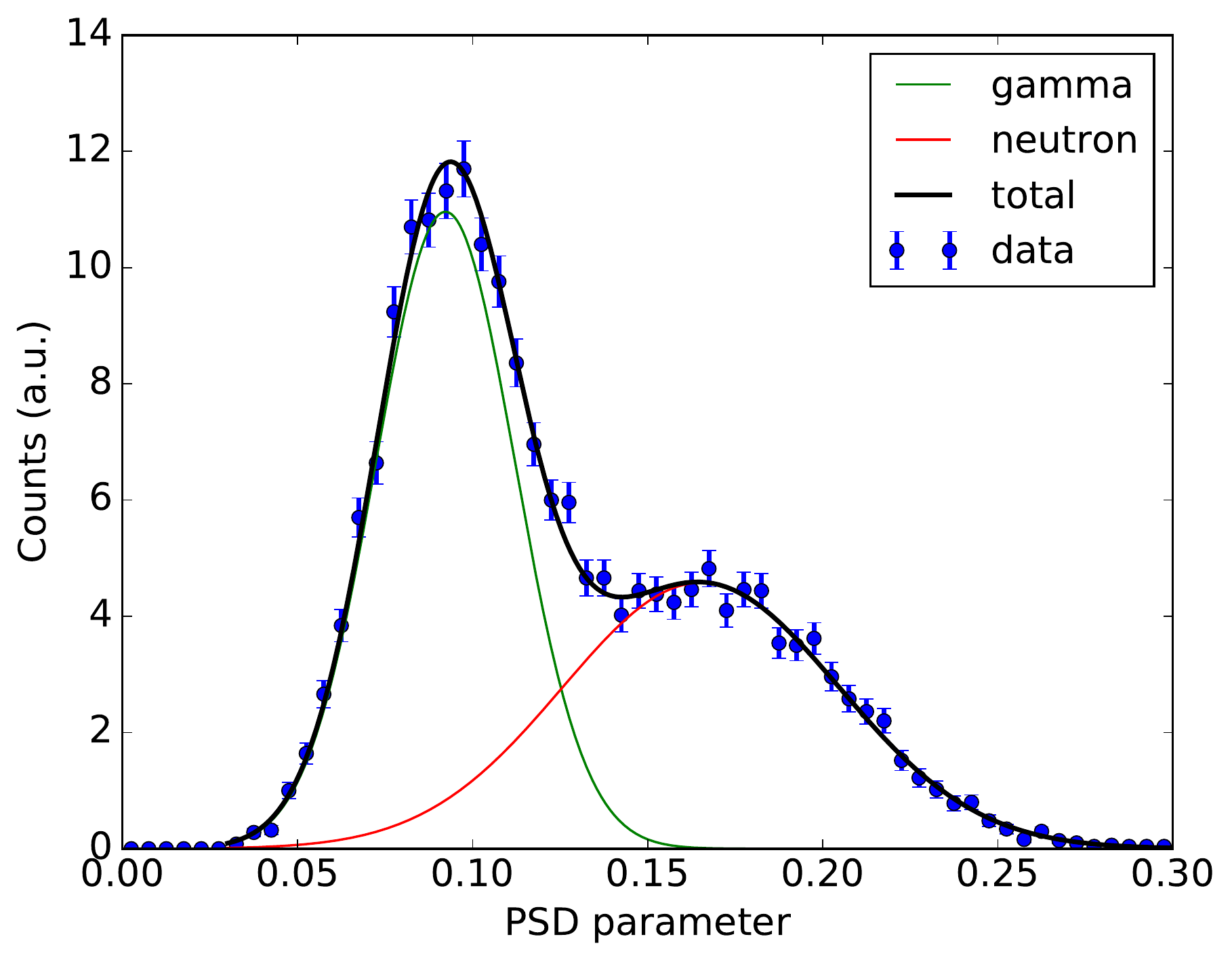}
\caption{The PSD distribution from cell 0, PMT 0 for an amplitude slice of 43099-43452 ADC counts, in which the contribution from 
the gamma (green) and neutron (red) Gaussians are shown.  The fit to the entire distribution, $f(x,E)$, is shown in black. }
\label{fig:fig6}
\end{figure}

\begin{figure}[tbp]\centering
\includegraphics[width=0.4\columnwidth]{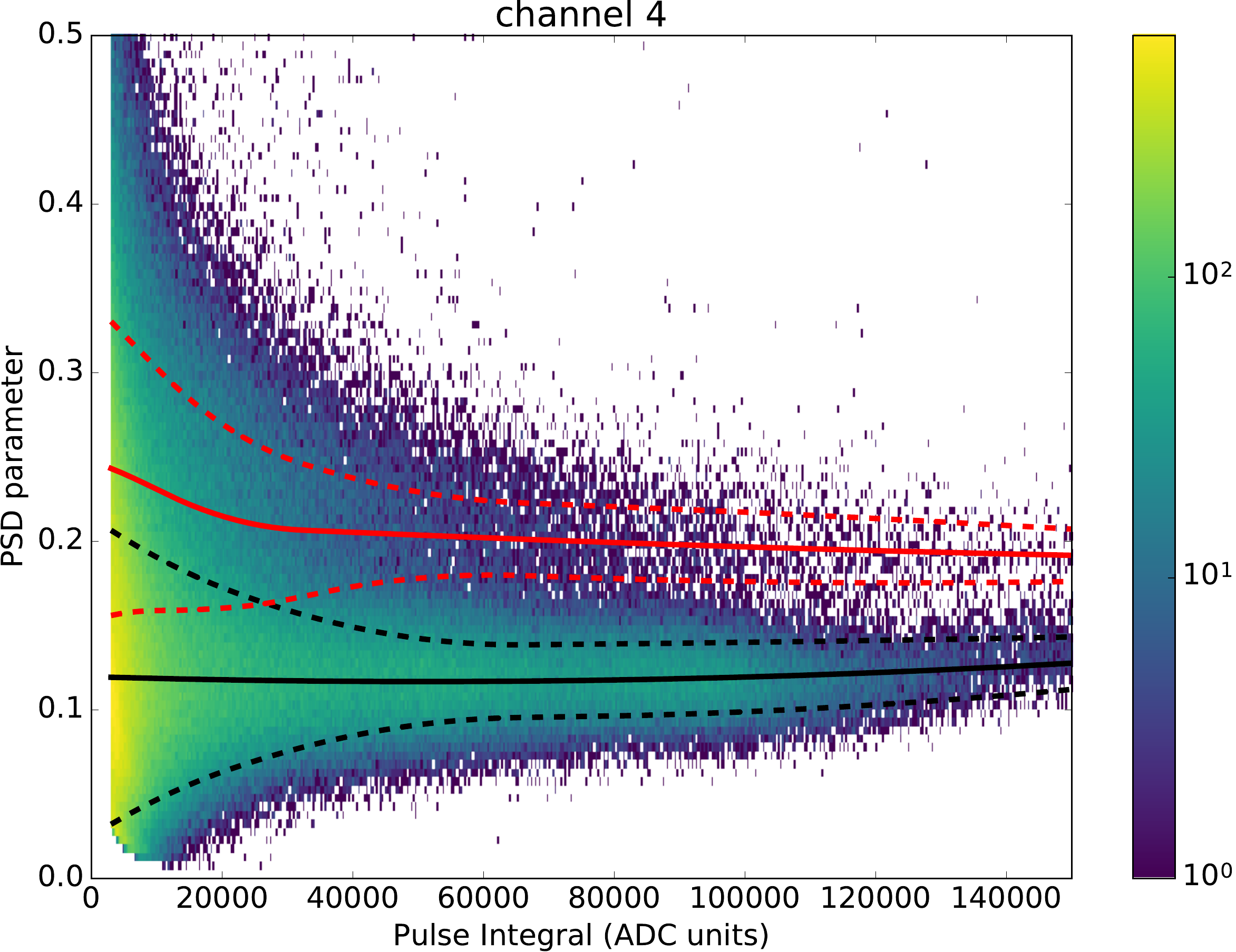}
\includegraphics[width=0.4\columnwidth]{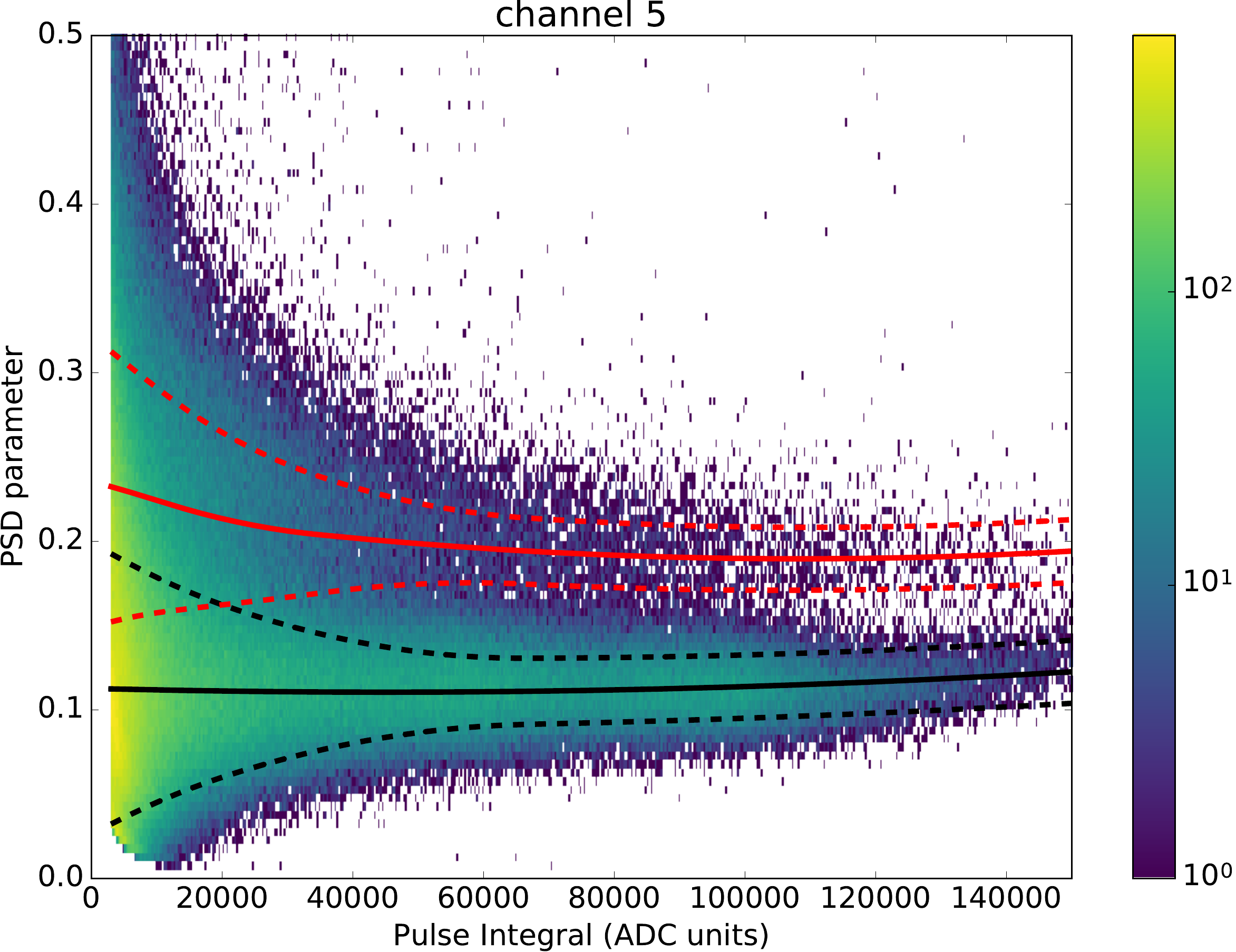}
\includegraphics[width=0.4\columnwidth]{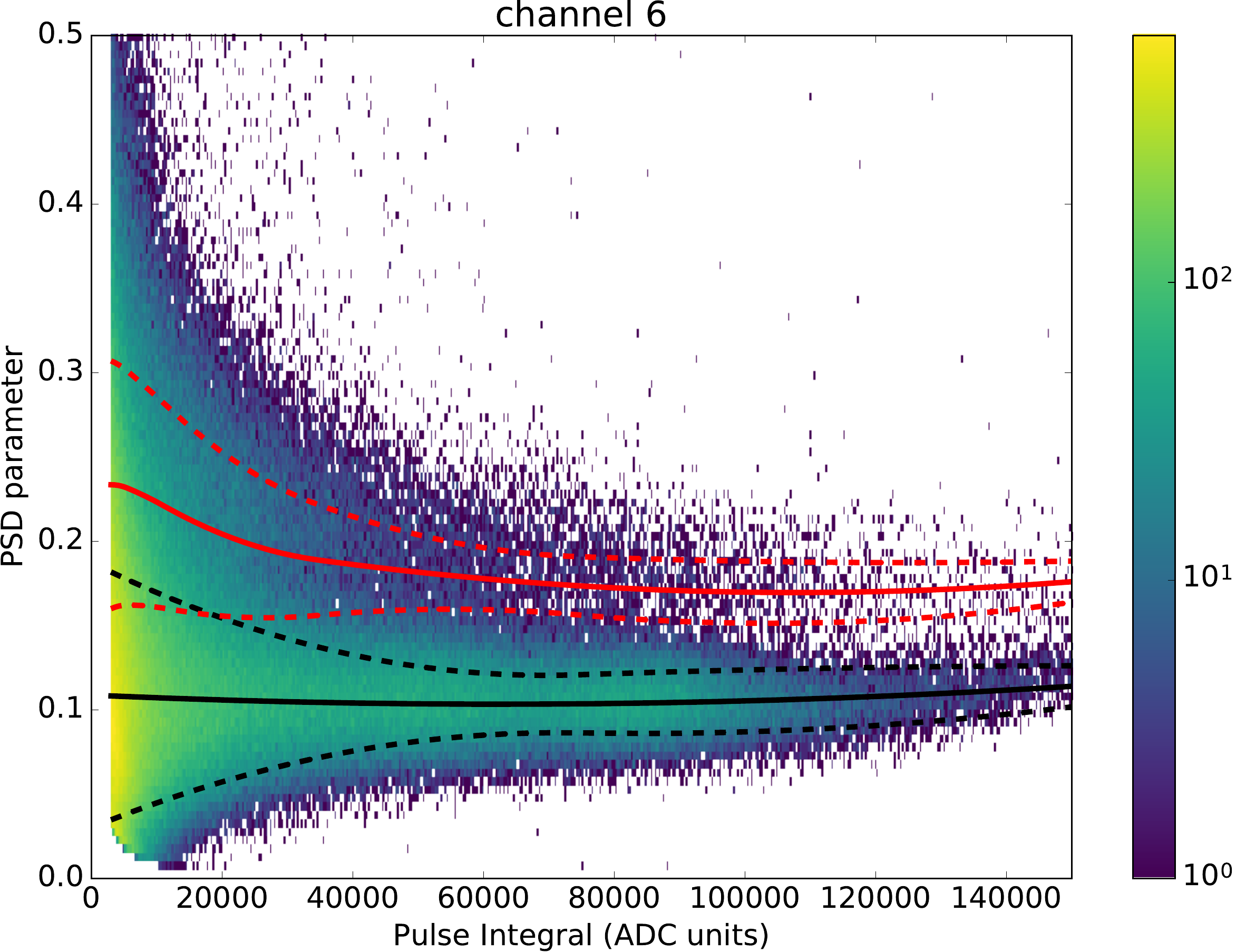}
\includegraphics[width=0.4\columnwidth]{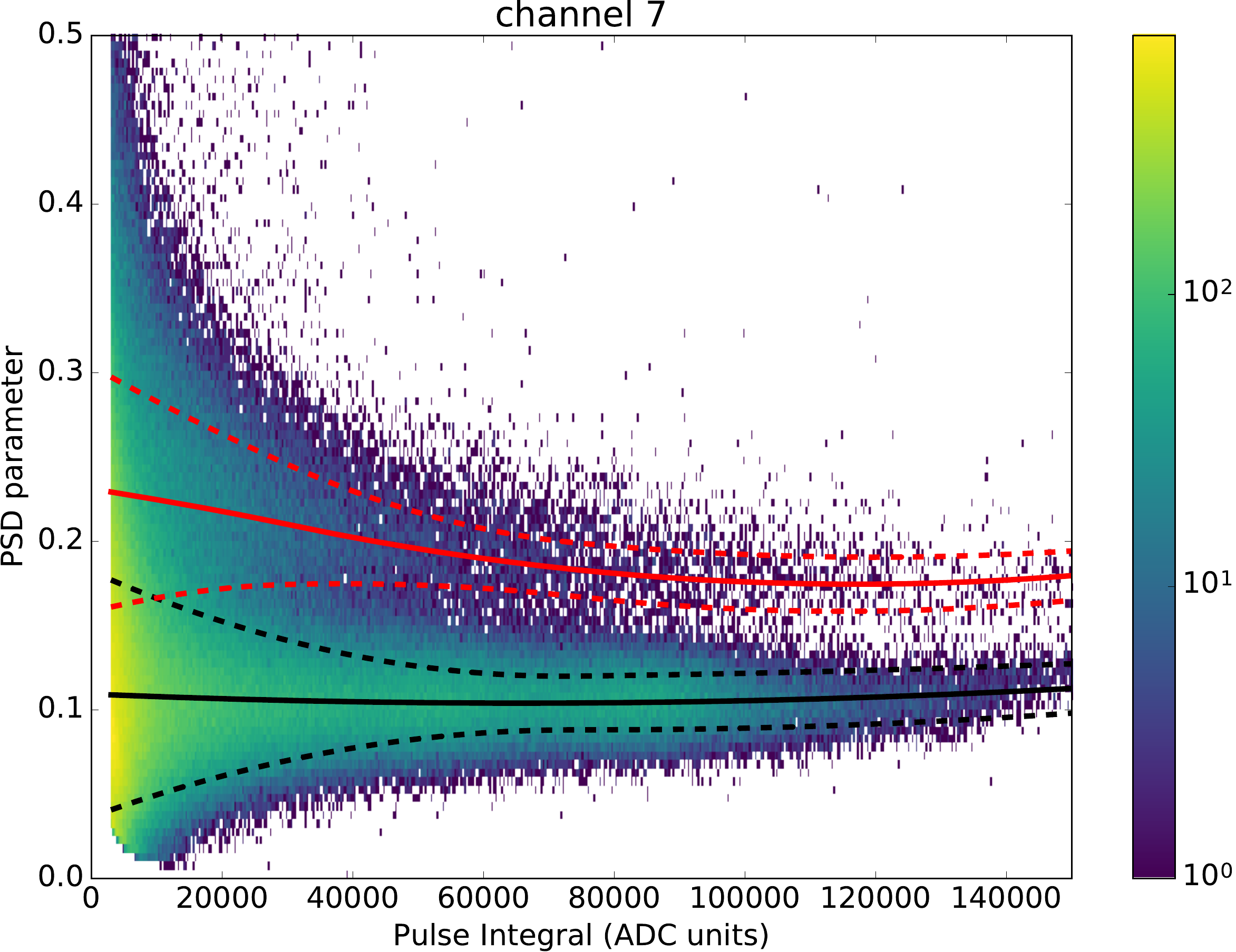}
\caption{All four PMTs within a cell contribute to the neutron probability defined in 
Equation \ref{eq:prob}.  The four individual PSD vs.\ energy distributions for cell 1 are 
shown here.  The mean and 1$\sigma$ bands are shown for both neutrons (red, and 
red-dotted) and gammas (black, and black-dotted).}
\label{fig:fig7}
\end{figure}

Because of the weak separation of the distributions, a gamma-pure dataset is used
to seed the parameters of the gamma band for the combined gamma/neutron dataset.
\Fig{fig:fig6} shows the PSD distribution for an amplitude slice in 
which the neutron and gamma bands are fairly well separated.  The neutron contribution 
to the distribution is indicated by the red curve, the gamma contribution by the green curve, 
and the total ($f(x,E)$) is the black curve.
\Fig{fig:fig7} shows the PSD parameter vs. energy for four PMTs in an 
individual cell.  The resulting mean and standard deviation from the fit slices are 
shown for both neutron and gamma populations: the neutron mean ($\mu_n$) is indicated by a 
red solid line, one standard deviation ($\mu_n \pm \sigma_n$) is indicated by red dotted lines, and the 
gamma mean ($\mu_\gamma$) and one standard deviation ($\mu_\gamma \pm \sigma_\gamma$) 
are similarly indicated by black solid and 
dotted lines.

We then define \Pn, the Bayesian probability that a given event is a neutron, as
\begin{linenomath}
\begin{equation}
\Pn = \frac{\LN}{\LN + R\LG + \LO},
\label{eq:prob}
\end{equation}
\end{linenomath}

where 
\begin{linenomath}
\begin{align*}
\LN=\prod_{p=0}^{4}\LNpE & ; & \LG=\prod_{p=0}^{4}\LGpE
\end{align*}
\end{linenomath}
are the product of the PSD value likelihoods for the individual PMTs, indexed by $p$.
The ``other'' likelihood value \LO is a constant factor added to allow for a third category in which the deposition is 
neither a gamma nor a neutron, such as a pileup event. Finally, the gamma-to-neutron ratio, $R$, is an energy-dependent factor 
which is determined for each event.
Double Gaussian fits (\eqnref{eq:dist}) over the PSD parameter distributions
are performed on the acquired dataset, fitting for $A_n$ and $A_{\gamma}$, but
fixing the $\mu$ and $\sigma$ values to those determined from the calibration
dataset. Then for a given event the gamma-to-neutron ratio is taken to be
the ratio of the average intensities of the gamma and neutron bands at the four
PMT amplitudes:
\begin{linenomath}
\begin{equation}
R = \frac{\sum_{p}{A_{\gamma,p}(E_p)}}{\sum_{p}{A_{n,p}}(E_p)}.
\end{equation}
\end{linenomath}

It should be noted that Equation \ref{eq:prob} defines a correct 
probability only under the conditions that the likelihoods are correctly represented by 
Gaussian distributions at all energies, that the values are independent and identically distributed,
that the ``other'' category is energy-independent and the value is correct, and finally that there are no other distributions 
contributing.  Some of these assumptions are known to be at best approximately true, 
so $\hat{P}_{n,\gamma}$ are estimates of the correct probabilities.

\subsection{Timing}

A mismatch between the encoder time (based on the CPU clock) and the digitizer
timestamps would result in an offset in the reconstructed source position 
compared to the true location. To avoid this, the encoder clock and digitizer are 
synced at the beginning of a run.  However, there remains a possibility of a drift in 
clocks over time, which can also result in positional alignment errors.  Significant drift
over a data run would cause phase shifts between rotations tending to flatten the 
counts vs. rotation angle distribution.  This is not observed, so any drift
between the clocks is small.

\section{Analysis}
\label{sec:ana}
Before source localization, all data is preprocessed.  In 
the preprocessing step a pulse shape parameter is extracted, and  then used to 
estimate the probability that each event is a neutron interaction based on 
the calibrations described above.  We require
the \Pn as defined in Equation \ref{eq:prob} to be greater than 0.999 
in order to be categorized as a neutron.
Each neutron event is then time-sorted and matched to a rotation angle based on the 
timestamp of the event and a time-tagged list of encoder positions.  The result of the 
preprocessing step is four arrays (or histograms) of neutron counts vs.\ rotation 
angle, one for each detector.  This is corrected for the detector live time as a function 
of rotation angle.  This correction is necessary due to uneven amounts of time spent 
accelerating.

Source detection and localization relies on calculating the likelihood of
observing a given dataset, both for a background-only hypothesis and for a
source-plus-background hypothesis. In this work, a simple isotropic background
model is used. For the source model, we assume the signal comes from a single
point source with a fission energy spectrum; since the location and source
strength cannot be known in advance, this is necessarily a composite model.
We contend that for most source search scenarios, a single-source model is
appropriate. If needed, however, both the signal and background models could
incorporate added complexity, including additional nuisance parameters
representing model unknowns.

Recall from \fig{fig:fig1} that $\zeta$ is the rotational angle of the
detector platform. An isotropic background model predicts an equal probability
for background events to be detected in each angle bin, resulting in a constant
background probability density function (pdf) $B(\zeta)$.
We bin the $\zeta$ space into $N_b$ bins, so the normalized pdf is:
\begin{linenomath}
\begin{equation}
B_d(\zeta) = 1/N_b.
\end{equation}
\end{linenomath}
The subscript $d$ indicates detector cell number, although for this background
model the four cells have identical pdfs.

\begin{figure}[htbp]
\centering
\includegraphics[width=0.8\columnwidth]{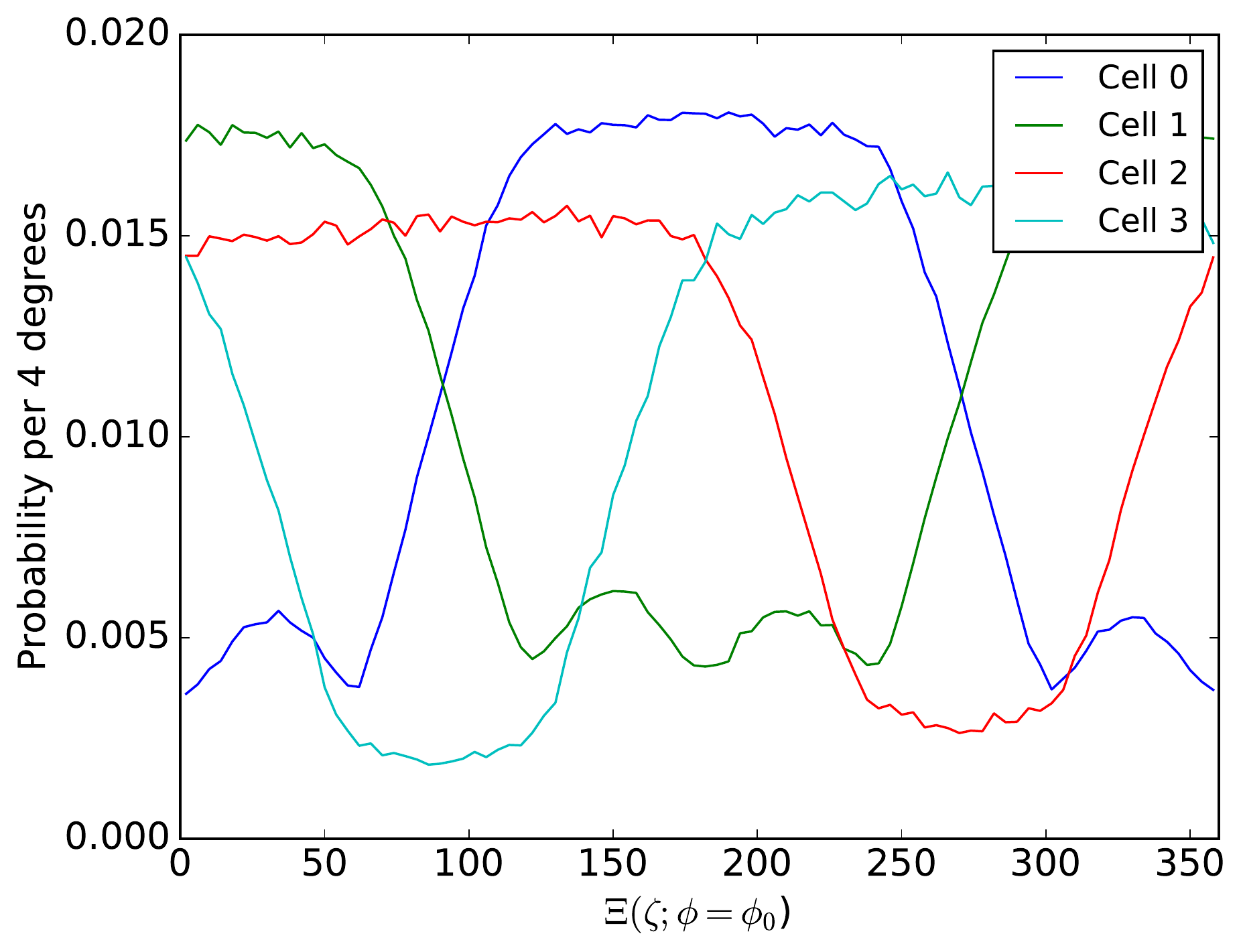}
\includegraphics[width=0.8\columnwidth]{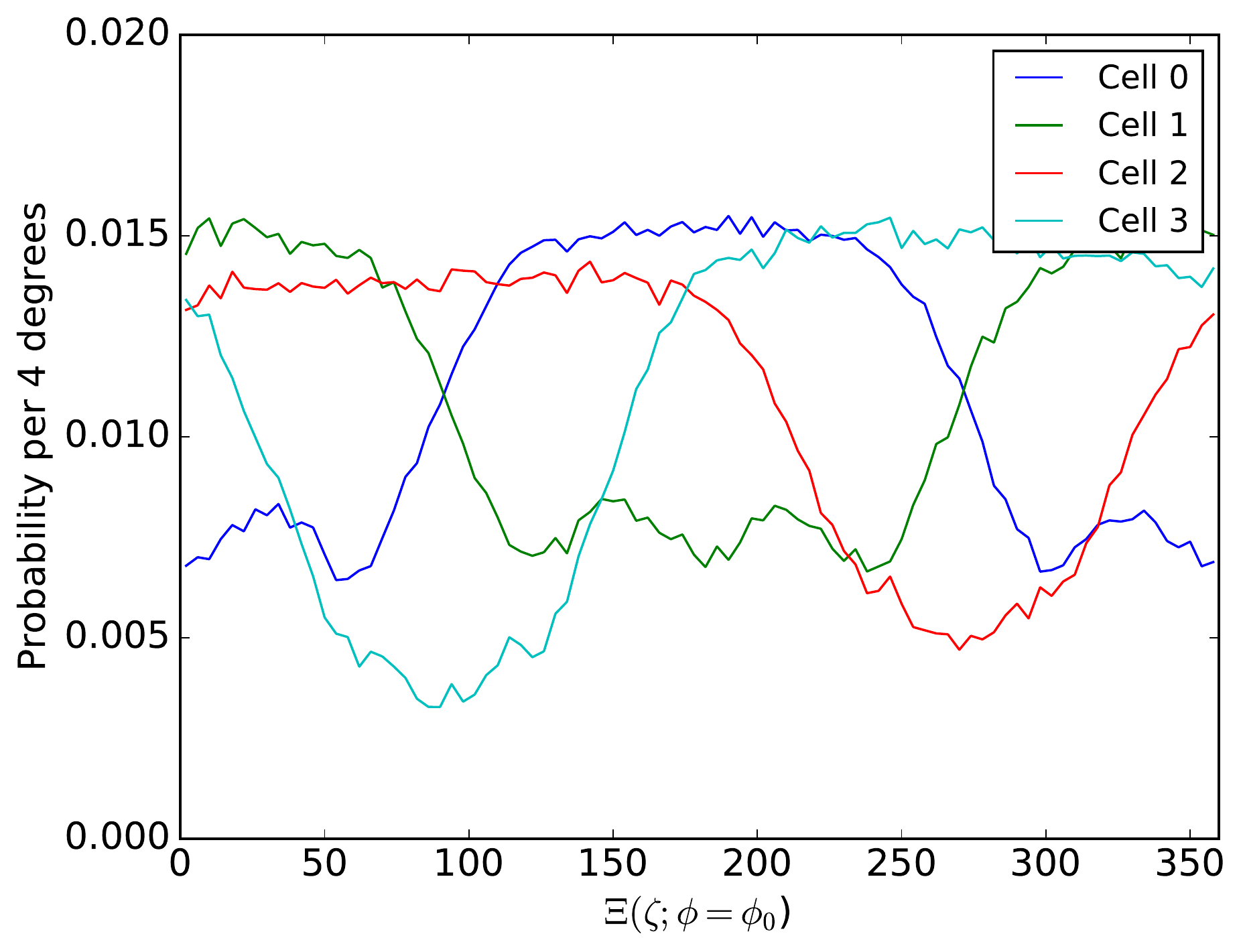}
\caption{The normalized and background subtracted neutron (top) and gamma (bottom) templates, i.e. $\Xi_d(\zeta;\phi$=$\phi_0$).}
\label{fig:fig8}
\end{figure}

The signal pdfs for each detector,
$\Xi_d(\zeta;\phi)$, where $\phi$ is the azimuthal angle of the
source location, are experimentally 
determined from a background-subtracted high-statistics measurement of a single 
strong neutron source positioned approximately \meters{10} away.
The measured templates are shown in \fig{fig:fig8}, top.
For reference, we also show the equivalent templates for a gamma selection
(\fig{fig:fig8}, bottom), which demonstrate less modulation as expected
since this system was designed for neutron sensitivity.
This measurement yields $\Xi_d(\zeta;\phi$=$\phi_0)$
for each of the four detectors, which is then 
phase-shifted through all possible source locations to determine the pdf for
any $\phi$.

Next we construct the expected number of counts $\lambda$ in angle bin $\zeta$,
given a source at location $\phi$. Since no \emph{a priori} constraints on the
signal strength or background rate are assumed, we introduce the signal fraction
\sigfrac and consider values in $0 \leq \sigfrac \leq 1$. 
The expected signal and background contributions are then calculated from $N_d$,
the total number of events observed in cell $d$, yielding  
\begin{linenomath}
\begin{equation}
\lambda_d(\zeta,\phi,\sigfrac) = N_{d}\left[\sigfrac\Xi_d(\zeta;\phi)
                               + (1-\sigfrac)B_d(\zeta)\right].
\end{equation}
\end{linenomath}

Finally, we can write the overall likelihood for a given dataset. The
background-only likelihood is determined by setting $\sigfrac=0$.
The measured data is represented by the number of events observed in each angle
bin for each detector cell, $n_d(\zeta)$.
(Note that $N_d = \sum_{\zeta}{n_d(\zeta)}$.)
The likelihood is taken to be the product of Poisson probabilities calculated
for each bin; in practice we calculate the log-likelihood:
\begin{linenomath}
\begin{equation}
\mathrm{log}\mathcal{L}(\phi,\sigfrac) =
   \sum_{d}\sum_{\zeta}\log\left(
	  \frac{e^{-\lambda_d(\zeta)}\lambda_d(\zeta)^{n_d(\zeta)}}{n_d(\zeta)!}
	 \right).
\label{eq:like}
\end{equation}
\end{linenomath}

\begin{figure}[tbp]\centering
\includegraphics[width=0.9\columnwidth]{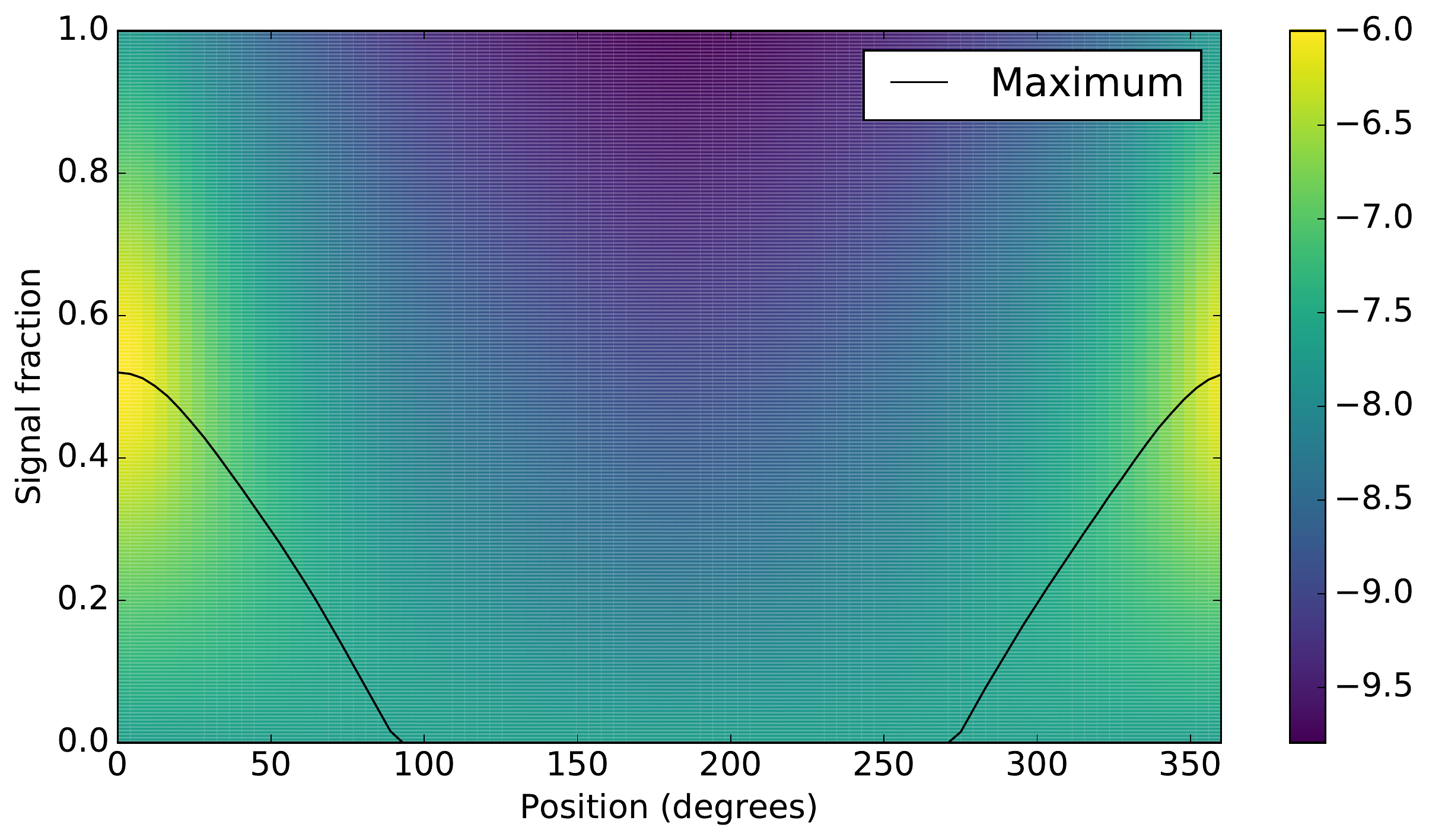}
\caption{The likelihood as a function of signal fraction \sigfrac and 
source position $\phi$.  The $z$ (color) axis is $-\log(-\log\mathcal{L})$, to make the behavior away from the maximum more visible.
The \sigfrac corresponding to the maximum likelihood at each source position is indicated by a solid black line.
In this example, the overall maximum likelihood is at $(\phi,\sigfrac) = (0,0.5)$.
}
\label{fig:fig9}
\end{figure}

To illustrate a typical result, \fig{fig:fig9} shows the likelihood as a function of $\phi$ and \sigfrac,
with a black line indicating the \sigfrac corresponding to the maximum likelihood for each $\phi$. The data were taken from a 24 hour measurement of a 21 $\mu$Ci $^{252}$Cf source at 10 m standoff. 
For this example, the overall maximum log-likelihood is at $(\phi,\sigfrac) = (0,0.5)$.
For $\sim 90 < \phi < \sim 270$, the maximum log-likelihood is at $\sigfrac=0$,
indicating that the background-only model is preferred over a source located at
any of those angles.

Using this log-likelihood, source detection is performed by constructing the
test statistic $D$, the maximum log likelihood ratio between the
source-plus-background and background-only hypotheses. The maximum is over all
possible source locations, $\phi$ from \degrees{0} to \degrees{360} in
\genunit{90}{steps}, and over all source strengths, \sigfrac from 0 to 1 in
\genunit{1000}{steps}:
\begin{linenomath}
\begin{equation}
D = \max_{\phi,\sigfrac}\left[-2\log\frac{\mathcal{L}(\phi,0)}
                        {\mathcal{L}(\phi,\sigfrac)}\right].
\label{eq:test}
\end{equation}
\end{linenomath}
The test statistic $D$ is then compared to a threshold to determine the
presence of a source. The threshold value to be used depends on the measurement
time and the desired tradeoff between source detection efficiency and false 
alarm rate.

Finally, for source localization we use $D(\phi)$, the log likelihood ratio
maximized over \sigfrac for a given $\phi$:
\begin{linenomath}
\begin{equation}
D(\phi) = \max_{\sigfrac}\left[-2\log\frac{\mathcal{L}(\phi,0)}
                                          {\mathcal{L}(\phi,\sigfrac)}\right].
\end{equation}
\end{linenomath}
Then the best estimate for the location of a putative source is
$\phihat=\operatorname{arg\,max}_\phi D(\phi)$.

\section{Measurements and Results}
\label{sec:data}
\begin{figure}[tbp]\centering
\includegraphics[width=0.99\columnwidth]{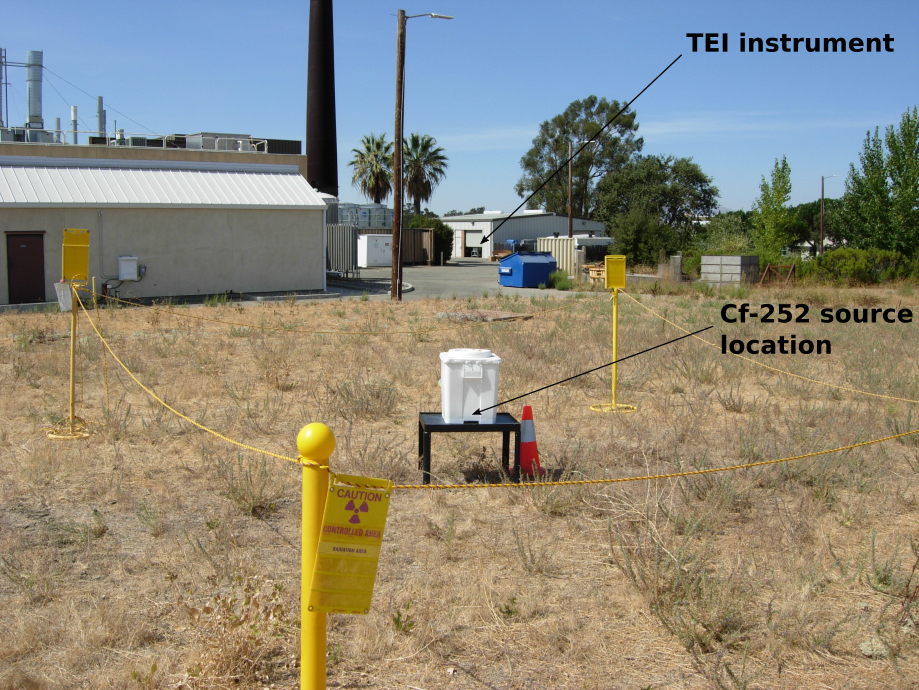}
\caption{Photo of the \Cftft source placed in a field at \meters{100} standoff from the detector.}
\label{fig:fig10}
\end{figure}

The measurements were taken with the 1D-TEI instrument kept at a stationary 
location and the source placed at increasing standoff locations of \meters{20},
\meters{40}, and 
\meters{100}.  The instrument was positioned inside a building and next to a large roll-up 
door.  A \Cftft source was placed inside a plastic container sitting on a metal 
table, with the source located \inch{17} off the ground.  The activity of the \Cftft source
was \genunit{1.03}{mCi}, known to 30\%, which corresponds to $4.4\pm1.3\times 10^6$ neutrons/s.  
A picture of the \Cftft source located at 
\meters{100} standoff from the instrument is shown in \fig{fig:fig10}.  

\begin{figure}\centering
\includegraphics[width=0.99\columnwidth]{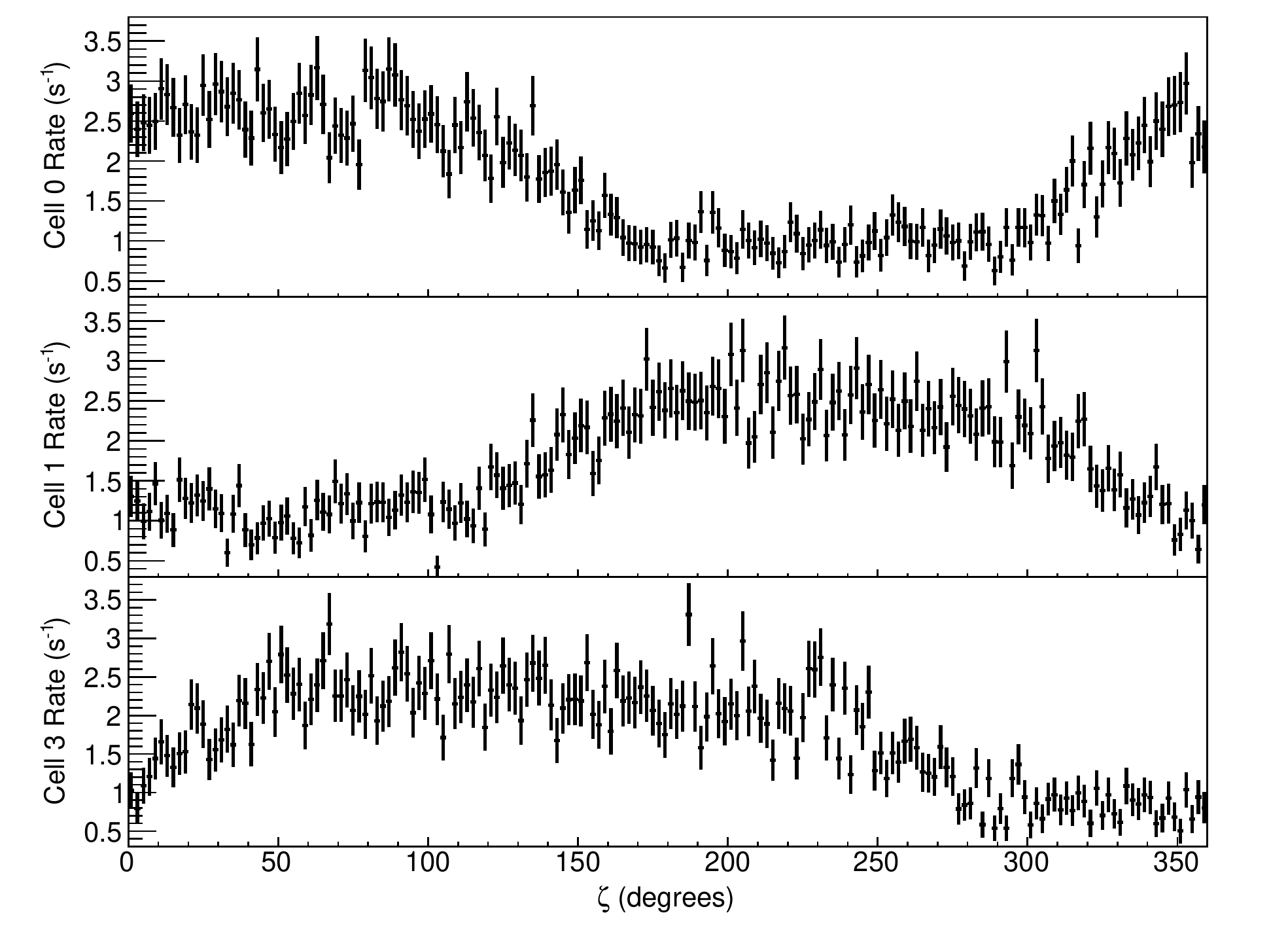}
\caption{Data for three working detector cells is shown for the \meters{40} standoff run.
Each plot shows the detected neutron rate in \degrees{2} bins of system rotation angle $\zeta$
for one of the cells.}
\label{fig:fig11}
\end{figure}

\begin{figure}\centering
\includegraphics[width=0.99\columnwidth]{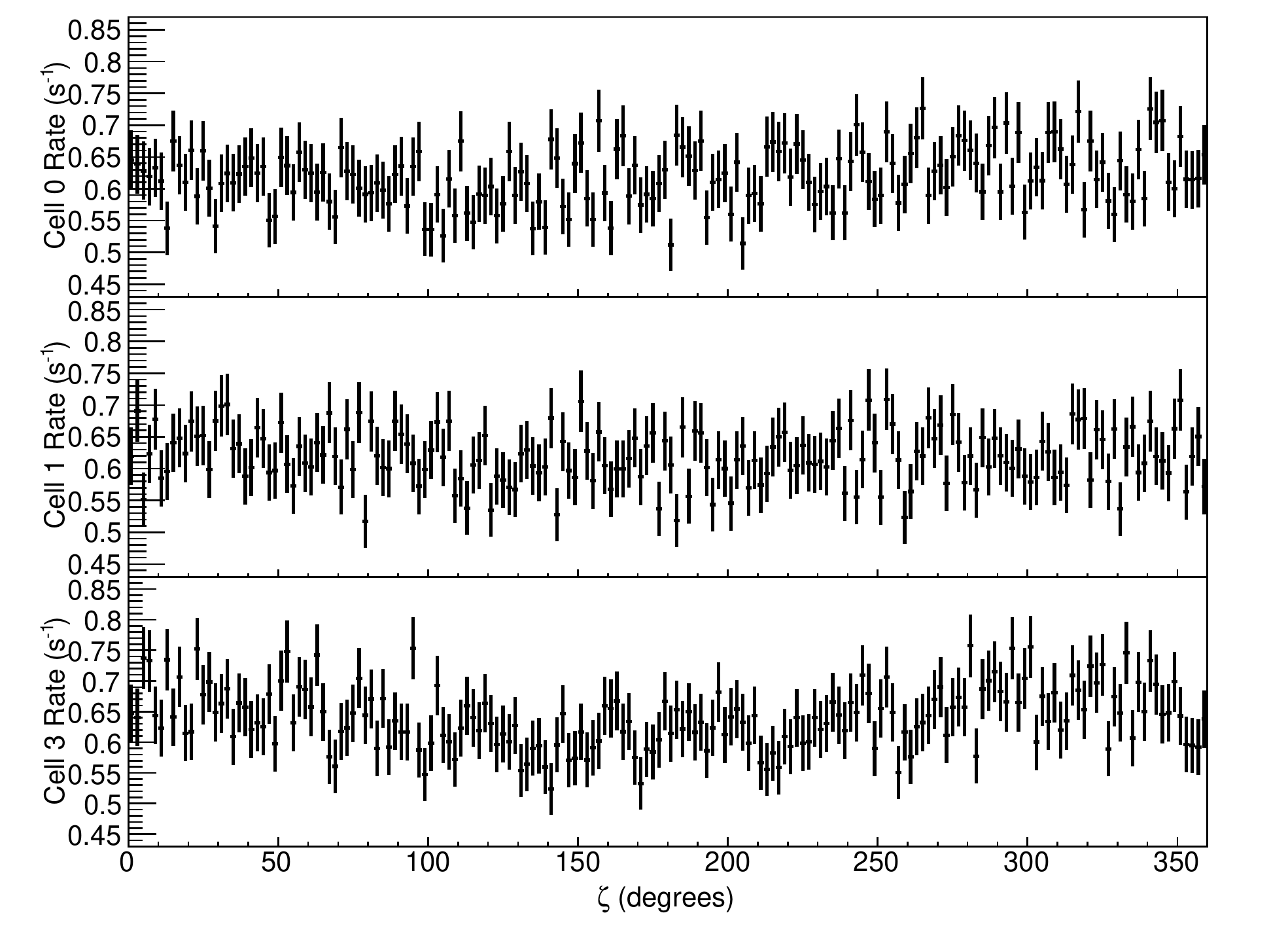}
\caption{Data for three working detector cells is shown for the background run.
Each plot shows the detected neutron rate in \degrees{2} bins of system rotation angle $\zeta$
for one of the cells.}
\label{fig:fig12}
\end{figure}

We acquired \minutes{10} of data for a \meters{20} standoff, \hours{1} for \meters{40}, \hours{4} for 
\meters{100}, and \hours{47} with no source present as a background dataset.
Care was taken to ensure that there was a clear line of sight between the 
source and the detector in order to minimize the effects of small-angle scatter.  
During these measurements, there was a malfunction in detector \#2,
one of the detectors on the short axis of the diamond; this detector was 
removed from the analysis, so all results below represent a three-detector system.
Data from the three working detectors for the \meters{40} standoff run is shown
in \fig{fig:fig11} after the preprocessing analysis stage.  For comparison, the data from a 
background run is shown in \fig{fig:fig12}.

\begin{figure}[tbp]\centering
\includegraphics[width=0.99\columnwidth]{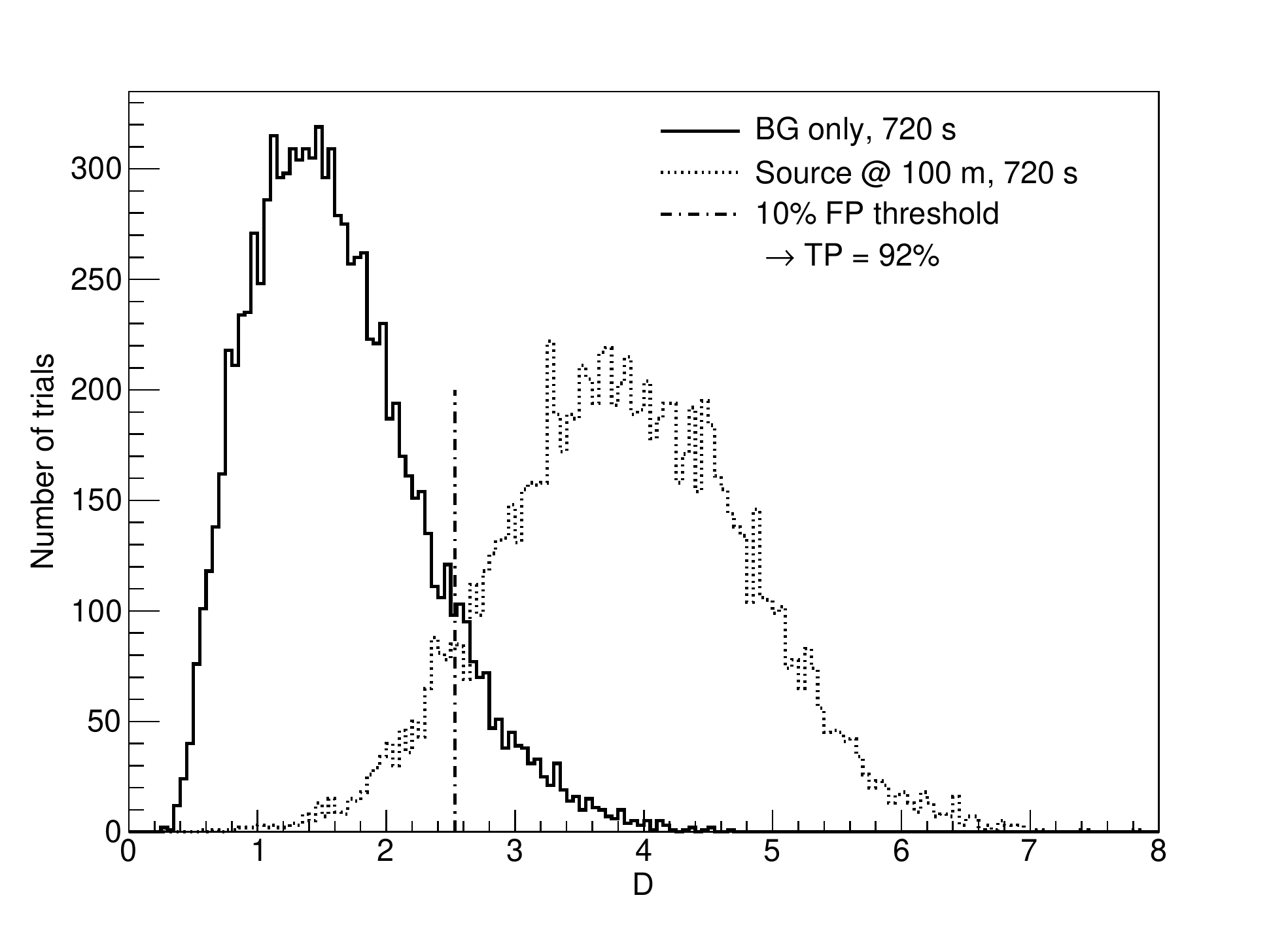}
\caption{
Histograms of the $D$ test statistic distribution for events sampled
from the background (solid) and \meters{100} signal (dotted) datasets. The number
of events corresponds to a \genunit{720}{s} measurement, which is the $t_d$
determined for \meters{100}. A line is drawn at the $D$ threshold that corresponds
to a 10\% false positive rate as determined from the background-only distribution.}
\label{fig:fig13}
\end{figure}

We calculate the test statistic $D$ according to Equation \ref{eq:test}.
For each standoff distance, we determine the dwell time that gives a 90\% detection efficiency with 
10\% false positives. This required time is denoted $t_{d}$ in 
order to distinguish it from the total run time $t_r$, and is determined as follows. For each distance,
we scan over a range of times $t$ and for each $t$ build a $D$ distribution
for the signal and background datasets. If $N$ events are observed in the
given dataset, 
$({t}/{t_r})\cdot N$ events are sampled with replacement and used to calculate $D$.
After repeating this sampling many times, a $D$ distribution is obtained. The
background $D$ distribution is used to set a 10\% false positive threshold (i.e.\
at the 90th percentile of the distribution); and the signal $D$ distribution is
used to calculate the source detection efficiency $\epsilon$ for this threshold.
The smallest time $t$ for which $\epsilon>90\%$ is taken as $t_d$.
For the \meters{100} dataset, \fig{fig:fig13} shows the background and signal
$D$ distributions for 10,000 sampled datasets at $t=t_d=\genunit{720}{s}$.

\begin{figure}[tbp]\centering
\includegraphics[width=0.99\columnwidth]{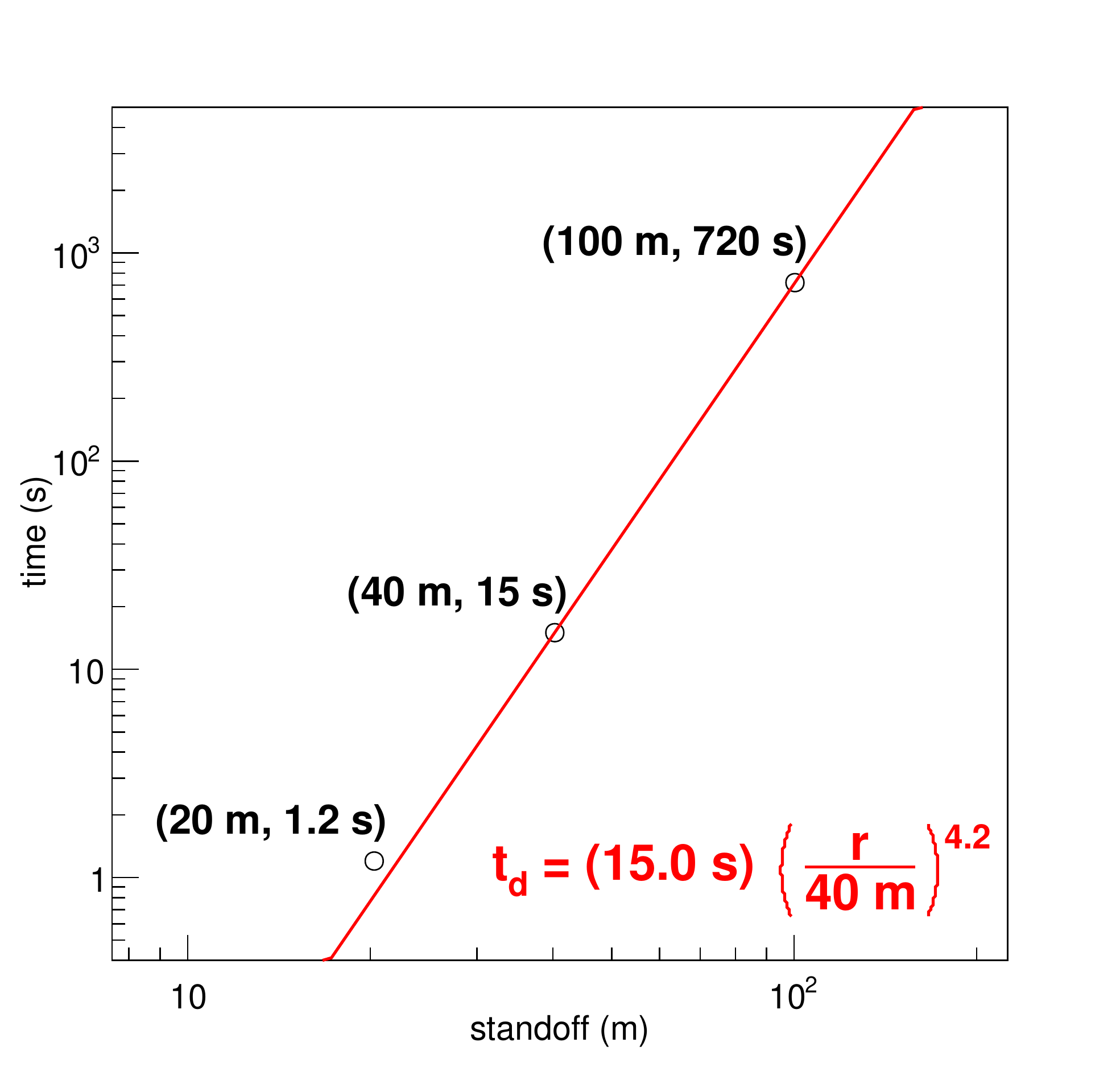}
\caption{The time $t_d$ yielding 90\% efficiency and 10\% false positives as a
function of standoff distance $r$ for measurements using a \genunit{1.03}{mCi}
($\pm30\%$) \Cftft source. The power law function fit result is shown in red.}
\label{fig:fig14}
\end{figure}

The time $t_d$ is plotted against standoff distance $r$ in \fig{fig:fig14};
also plotted is a fit to a power law 
($t_d=p_0 r^{p_1}$), which yields an exponential power of 4.2.
Indeed, $t_d \propto r^4$ is expected behavior from analysis
on the simple case of significance in a counting experiment:
$S=\NS/\sqrt{\NB}$, where \NS and \NB are the number of signal and background
counts, respectively. Since both \NS and \NB scale with acquisition time
$t$, we see that $S\propto\sqrt{t}$. Changing the distance to the source
affects the flux at the detector via $1/r^2$, which directly affects the
significance, $S\propto 1/r^2$. Thus we see that to maintain a constant
significance $S$ under changes in $r$, we need $t\propto r^4$.

\begin{figure}[tbp]\centering
\includegraphics[width=0.99\columnwidth]{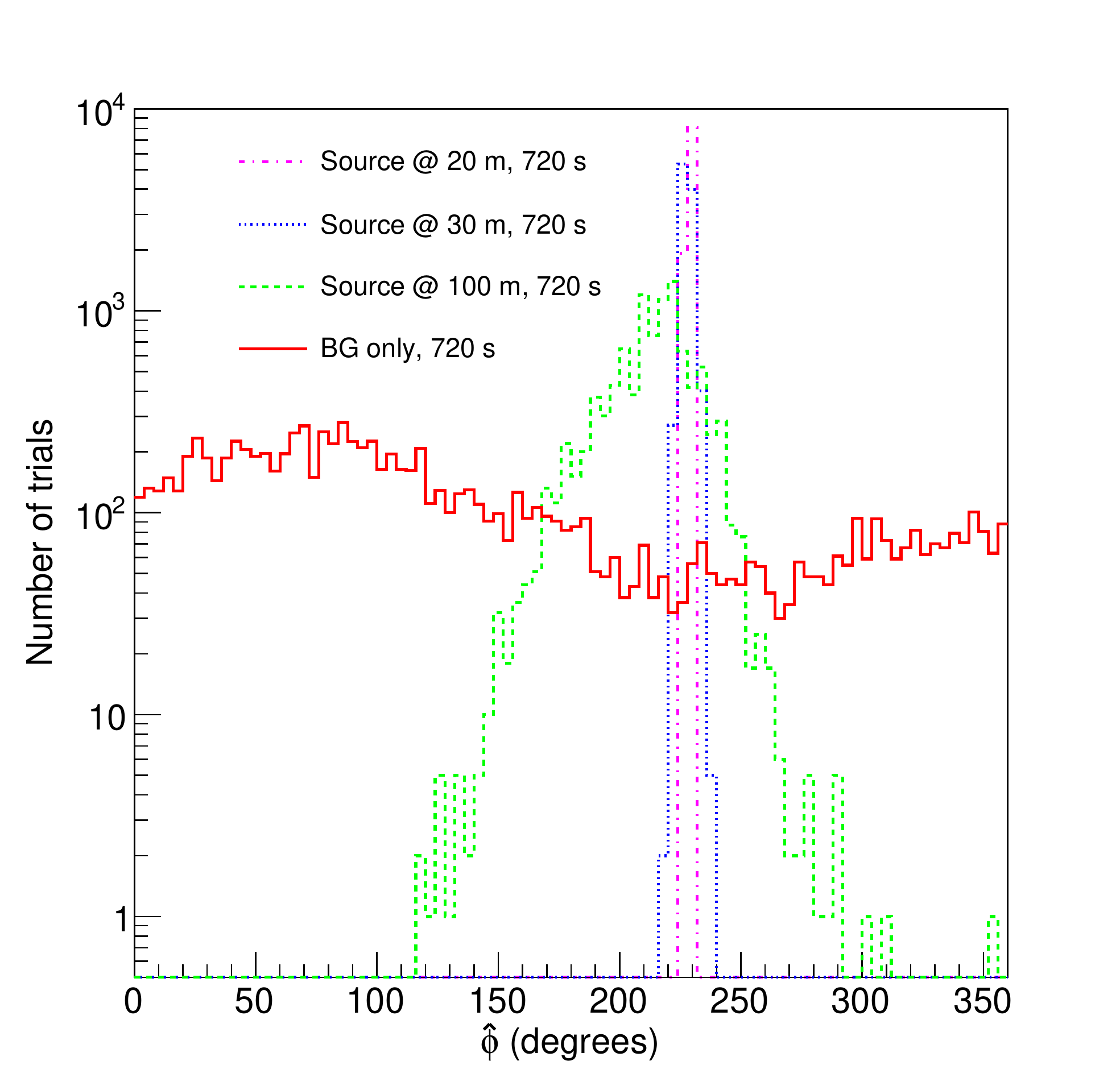}
\caption{The localization performance is evaluated by plotting \phihat, the estimated source azimuthal angle, for ensembles of resampled datasets corresponding to $t=\genunit{720}{s}$. The true source location for the runs with source present is not precisely known and is not necessarily identical for the three standoff distances, however the peak of each distribution is approximately located in the expected position.}
\label{fig:fig15}
\end{figure}

To evaluate the localization resolution of the system, we sample with replacement as described above to create 10,000 resampled datasets for each source standoff distance and for the background-only dataset. We use $t=\genunit{720}{s}$, i.e.\ $t_d$ as determined for the \meters{100} source
standoff. In \fig{fig:fig15}, we plot the distributions of \phihat, the estimated source azimuthal angle, for each ensemble of resampled datasets. The true source location for the runs with source present is not precisely known (and is not necessarily identical for the three standoff distances), so we do not evaluate a bias in the estimated direction. The RMS of the \phihat distribution is \degrees{1.6}, \degrees{2.5}, and \degrees{19} for \meters{20}, \meters{40}, and \meters{100} standoff, respectively, providing an estimate of the system's angular resolution as a function of number of signal events and S:B ratio. We observe some non-uniformity in the background-only dataset. It peaks in the opposite direction from the source position in the field, which may indicate that the non-uniformity is due to neutron production (i.e.\ the ``ship effect'') on large equipment in the building. Note that the likely effect of this particular background non-uniformity on the runs with source present is to reduce the significance of the rate modulation and increase $t_d$. 

The analyzed results for $t_d$ and \phihat RMS are summarized in \tab{t:results}, along with the measured neutron rates and inferred S:B for each run.

\begin{table}\centering
\caption{Summary of conditions and results for the data runs presented in this article. The signal-to-background ratio S:B is estimated using the measured rates. See text for details on other quantities.}
\label{t:results}
\begin{tabular}{rrrrr}
\hline
\thead{Standoff \\ distance (m)} & \thead{Overall \\ rate (s$^{-1}$)} & \thead{S:B} & \thead{$t_d$ (s)} & \thead{\phihat RMS \\ (deg)} \\
\hline
20  & 25.2 & 12   & 1.2 & 1.6 \\
40  & 5.3  & 1.8  & 15  & 2.5 \\
100 & 2.1  & 0.14 & 720 & 19  \\
BG  & 1.9  & N/A  & N/A & N/A \\
\hline
\end{tabular}
\end{table}
	
\section{Conclusions}
We have presented the design, characterization, and results for a prototype radiological search 
system.   We have demonstrated the ability to detect a \genunit{\sim 1}{mCi} \Cftft radiological 
source at \meters{100} stand off with 90\% detection efficiency and 10\% false positives 
against background in \minutes{12}.  This same detection efficiency is met at \genunit{15}{s}
for a \meters{40} standoff, and \genunit{1.2}{s} for a \meters{20} standoff.

\section*{Acknowledgments}
Sandia National Laboratories is a multimission laboratory managed and operated by National Technology and Engineering Solutions of Sandia, LLC, a wholly owned subsidiary of Honeywell International, Inc., for the U.S. Department of Energy's National Nuclear Security Administration under contract DE-NA0003525.

This material is based upon work supported by the U.S. Department of Homeland 
Security under Grant Award Number, 2012-DN-130-NF0001. The views and 
conclusions contained in this document are those of the authors and should not be 
interpreted as representing the official policies, either expressed or implied, of the 
U.S. Department of Homeland Security.

We would like to thank the US DOE National Nuclear Security Administration,
Office of Defense Nuclear Non-proliferation for funding this work.

\end{document}